\documentclass[citeautoscript,floatfix,aps,prl,twocolumn,%
superscriptaddress,longbibliography,amsmath,amssymb,10pt]{revtex4-2}

\usepackage{graphicx,bm,pdfpages,pgffor}

\makeatletter
\AtBeginDocument{\let\LS@rot\@undefined}
\makeatother

\begin{document}

\title{Tilt-driven antiferroelectricity in PbZrO$_3$}
\author{K. Shapovalov}
\affiliation{Institut de Ci\`encia de Materials de Barcelona (ICMAB-CSIC), Campus UAB, 08193 Bellaterra, Spain}

\author{M. Stengel}
\affiliation{ICREA --- Instituci\'o Catalana de Recerca i Estudis Avan\c cats, 08010 Barcelona, Spain}
\affiliation{Institut de Ci\`encia de Materials de Barcelona (ICMAB-CSIC), Campus UAB, 08193 Bellaterra, Spain}

\begin{abstract}
  Antiferroelectricity is a state of matter that has so far eluded a clear-cut definition.
  Even in the best-known material realization, PbZrO$_3$, the physical nature of the driving force towards an antipolar order has not been settled yet.
  Here, by building a Landau-like continuum Hamiltonian from first-principles via an exact long-wave approach, we reconcile the existing theories in terms of a single physical mechanism.
  In particular, we find that a formerly overlooked trilinear coupling between tilts, tilt gradients and polarization provides a surprisingly accurate description of the energetics and structure of the antiferroelectric ground state of PbZrO$_3$.
  We discuss the relevance of our findings to other ferrielectric and incommensurate polar structures that were recently observed in perovskites.
\end{abstract}

\maketitle

\begin{figure}
  \centering
  \includegraphics[width=\linewidth]{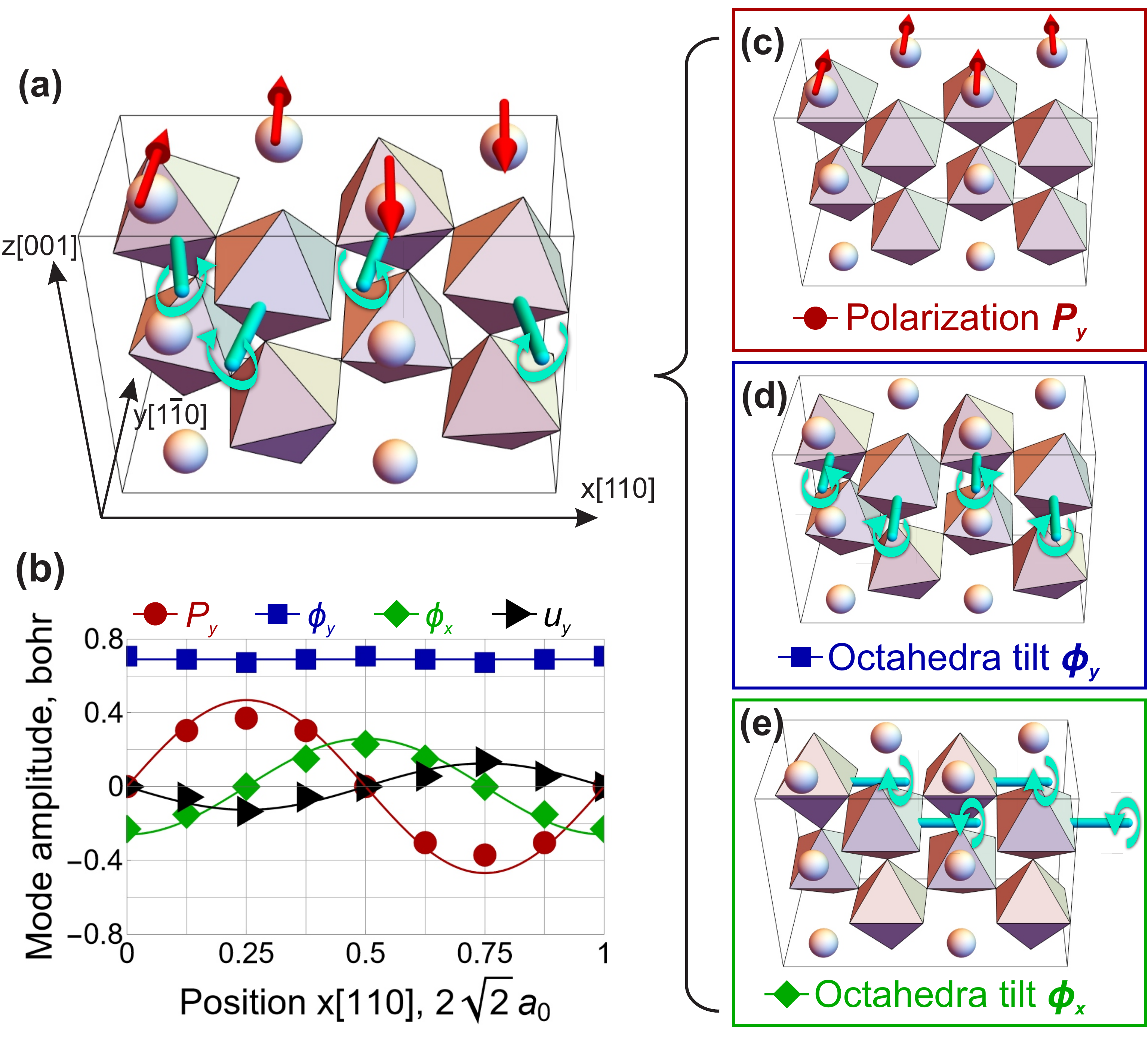}
  \caption{(a) Calculated atomic structure of the $Pbam$ phase of PZO. Only Pb atoms (spheres) and O$_6$ octahedra are shown. Red arrows indicate polar displacements of Pb atoms, blue arrows indicate ${\rm O}_6$ octahedra tilts. (b) Decomposition of the atomic distortions shown in (a) over polarization $P_y$, AFD octahedral tilts $\phi_x$, $\phi_y$, and displacement field $u_y$.
  Symbols and solid lines correspond to the layer-by-layer and Fourier decompositions, respectively. 1~bohr of amplitude corresponds to 174~$\mu\rm C/\rm{cm}^2$ of polarization and $14.8^\circ$ of octahedral tilt. (c--e) Atomic distortions described by the three primary order parameters
  ($P_y$, $\phi_y$, $\phi_x$).} \label{Fig:AFE-Decompose}
\end{figure}

Antiferroelectric (AFE) materials have received increasing attention recently,
both for their potential applications (e.g. in energy storage) and their fundamental interest \cite{Rabe13Chapter,Hao14,Wei14,Xu17}.
PbZrO$_3$ (PZO) is by far the best studied example, and yet its theoretical
understanding has so far proven to be elusive \cite{Tagantsev13,Iniguez14,Hlinka14,Xu19,Fu20,Aramberri21}.
During its AFE phase transition at $\sim500$~K, PZO changes its crystalline symmetry
from cubic $Pm\bar3m$ to orthorhombic $Pbam$, with Pb ions forming the
$\uparrow\uparrow\downarrow\downarrow$ pattern periodically repeated in a $\langle110\rangle$ direction (Fig.~\ref{Fig:AFE-Decompose}(a)).
This phase transition is accompanied by additional non-trivial atomic distortions, including antiferrodistortive (AFD) O$_6$ octahedra tilts \cite{Tagantsev13,Iniguez14,Xu19}, whose relationship to the antipolar pattern is not well understood.
Several models have been proposed to explain the physical origin of antiferroelectricity in PZO.

In the first major model, Tagantsev {\em et al.} \cite{Tagantsev13} rationalize AFE as a manifestation of an incommensurate (IC) phase transition driven by flexoelectricity.
Flexoelectricity, describing the coupling between electrical polarization
and strain gradients \cite{Zubko13,Yudin13,Stengel16chapter,Wang19}, enters the
continuum bulk Hamiltonian via a Lifshitz invariant, which can stabilize modulated
phases if the coupling is sufficiently strong \cite{Axe70,Tagantsev13,Pottker16}.
If, moreover, the optimal period of the modulation is close to a multiple of the
cell parameter, it can spontaneously lock-in to a commensurate phase \cite{Bak82}, which would
explain \cite{Tagantsev13} the $\uparrow\uparrow\downarrow\downarrow$ AFE pattern of Pb displacements in PZO.
This idea appears to be very reasonable: IC phases have indeed been reported in PZO \cite{Burkovsky17} and closely related perovskites \cite{Fu20,Bosak20,Ma19}.
However, there are conflicting experimental reports on whether flexoelectricity is strong enough for such a mechanism to be viable \cite{Tagantsev13,ValesCastro18}.
Moreover, flexoelectricity alone does not explain the simultaneous condensation
of the AFD octahedral tilt mode, whose amplitude and contribution to the energetics are remarkably large in the ground-state structure \cite{Iniguez14,Burkovsky19}.

The second model, proposed by \'I\~niguez and coworkers \cite{Iniguez14}, invokes a trilinear coupling between the
antipolar displacement of the Pb atoms
[a $\Sigma$-point distortion with wavevector $\mathbf{q_\Sigma}=(1/4,1/4,0)$],
the AFD octahedra tilt mode [$R$-point, $\mathbf{q_R}=(1/2,1/2,1/2)$]
and a third $S$-point mode [$\mathbf{q_S}=(1/4,1/4,1/2)$].
Such $\Sigma$--$R$--$S$ interaction would lower the energy when the three modes coexist,
and hence explain their simultaneous condensation.
Their first-principles-based parametrization confirms that this coupling is indeed
essential for the $Pbam$ structure to win over competing low-energy phases \cite{Iniguez14}.
The main conceptual issue is that $\Sigma$- or $S$-phonons can hardly be regarded as
``elementary fields'' in the theoretical study of perovskite materials -- both of them correspond
to complex distortion patterns, whose physical interpretation is (especially in
the case of the $S$-mode) unclear.
In other words, these modes are specific to the ground-state structure of
PZO, so their physical properties appear difficult to generalize to other
systems where the antipolar ordering occurs with different symmetries and/or
periodicities \cite{Burkovsky19,Fu20}.

Here we develop a unified theoretical framework that merges the two models into a single physical mechanism.
In particular, we show that the transition is driven by a trilinear coupling term that is
akin to flexoelectricity, but involves gradients of the AFD octahedral tilt modes, rather than of the strain.
Such coupling is ubiquitous in the physics of (multi)ferroics, its
effects ranging from the generation of polarity at the twin boundaries in SrTiO$_3$ \cite{Schiaffino17} to polar textures in inversion-symmetry-broken magnets \cite{Mostovoy06}; this and similar mechanisms have been discussed in the context of IC/AFE phases \cite{Eliseev13,Hlinka14}.
Thus, our interpretation saves the idea of an incommensurate phase transition, while endowing the $S$-mode with a clear physical interpretation as a modulated tilt mode.
By establishing an exact mapping between continuum theory and first-principles simulations in the long-wavelength limit, we validate the continuum approach, qualitatively and quantitatively reproducing the energetics and structure of the ground-state AFE structure within our model.
Our results provide the missing link required for triggering the IC phase transition in PZO and
related materials, and can be regarded as a universal pathway towards incommensuration in ferroics.

{\em Structural analysis.} In order to discuss the emergence of modulated phases, we need first of
all to establish a rigorous 
mapping between spatially inhomogeneous order parameters and the 
atomic structure of PZO.
Following earlier works \cite{Schiaffino17,Stengel16}, we identify the order parameters
with the eigenvectors of the force-constant matrix calculated {\em ab initio} at high-symmetry points of the parent $Pm\bar3m$ Brillouin zone [see Supplemental Material 1 \cite{SM} for the details of the density functional theory (DFT) calculations].
More specifically, we consider the following two types of distortions: (i) zone-center modes, including the ``soft'' polar mode ($\bf P$) and the acoustic branch ($\bf u$), and (ii) out-of-phase AFD tilts of the oxygen octahedra ($\bm\phi$), located at the $R$-point of the cubic phase.
To extract the continuum fields from a given atomic distortion pattern, we then operate either in Fourier space [via a decomposition into $\bf q$-points of the $Pm\bar3m$ Brillouin zone and a subsequent projection onto the basis (i--ii)], or in real space [by constructing, starting from the cell-periodic patterns (i--ii), a space-resolved basis of ``local modes'' centered on $(110)$ layers \cite{SM}].
In the following we shall test both approaches on the first-principles-calculated $Pbam$ phase of PZO [Fig.~\ref{Fig:AFE-Decompose}(a)] and check their mutual consistency.

Our Fourier decomposition shows, in agreement with previous studies \cite{Iniguez14,Xu19},
that the ${\bf q_R}$, ${\bf q_\Sigma}$  and ${\bf q_S}$ modes of the $Pbam$ phase account,
respectively, for 63.0\%, 33.2\% and 3.8\% of the total distortion.
In Fig.~\ref{Fig:AFE-Decompose}(b) we show the spatially resolved order parameters, plotted as functions of the $x$ coordinate.
[$x$ corresponds to the $[110]$ modulation direction, $y$ to the antipolar displacement of the Pb atoms
along $[1\bar10]$, see Fig.~\ref{Fig:AFE-Decompose}(a)].
Both layer-by-layer and Fourier approaches result in similar field amplitudes, confirming that our analysis is physically sound.
The largest distortion corresponds to a uniform AFD mode with the tilt axis oriented along $y$,
$\phi_y$ [Fig.~\ref{Fig:AFE-Decompose}(d)], accounting for the near totality (99.8\%) of the $R$-mode.
Next, we identify a sinusoindal modulation of the polarization, $P_y$ [Fig.~\ref{Fig:AFE-Decompose}(c)],
which reflects the characteristic $\uparrow\uparrow\downarrow\downarrow$ antiferroelectric
displacement of the Pb ions and constitutes the largest (92.0\%) contribution
to the $\Sigma$-mode.
The remainder of the $\Sigma$ distortion is predominantly due to the
acoustic mode, $u_y$, whose amplitude describes the local displacement of the unit
cell along $y$.
Finally, and most importantly, our analysis reveals a \emph{secondary} tilt mode, $\phi_x$
[Fig.~\ref{Fig:AFE-Decompose}(e)], which has not been reported before.
This mode has the tilt axis oriented along $x$ (longitudinal), and is modulated with the
same period as the polarization, its phase shifted by 90$^\circ$ [Fig.~\ref{Fig:AFE-Decompose}(b)].
A closer look shows that this modulated tilt coincides to a high accuracy (92.9\%) with
the $S$-mode that was reported in the literature \cite{Iniguez14}.
Together, the four modes plotted in Fig.~\ref{Fig:AFE-Decompose}(b) amount to 99.0\% of the total distortion amplitude.

The above analysis unambiguously identifies the main distortions
of the $Pbam$ phase, including the ``exotic'' $\Sigma$ and $S$
modes of the standard decomposition, as spatial modulations of polar and tilt modes.
This constitutes a drastic conceptual simplification, since it allows us to describe a
structure as complex as the AFE ground state of PZO in terms of
``elementary'' fields that are ubiquitous in perovskite materials.
It presents practical advantages, too: in combination with recent developments in first-principles
theory~\cite{Stengel13,Stengel16chapter,Stengel16,Schiaffino17,Royo19} that enable systematic calculation of
gradient-mediated couplings, the 
aforementioned mapping
between lattice modes and continuum fields
allows for a quantitative (i.e., free of
fitted parameters or phenomenological assumptions) validation of the physical
mechanisms that were proposed so far as driving force towards the antiferroelectric state.

{\em Flexoelectricity.} In general, the gradient coupling between a strain $\varepsilon$ and a polar mode $P$ can be written as
\begin{equation}
  E_{\rm fxe}=-fP\frac{\partial\varepsilon}{\partial x},
\end{equation}
where $f$ is the flexoelectric coupling coefficient \cite{Yudin13,Wang19,Stengel16}.
This expression is typically discussed in conjunction with the standard Landau-Ginzburg-Devonshire
free energy of a ferroelectric, which in its simplest form contains
the homogeneous Landau potential, $E_{\rm hom}(P) = AP^2 / 2 + BP^4 /4$,
the correlation energy, $E_{\rm corr}(P) = G  (\partial P/\partial x)^2 / 2$, and the
elastic energy $E_{\rm elas}(\varepsilon) = C \varepsilon^2 / 2$ \cite{Yudin13}.
A strong flexoelectric coupling may trigger a transition to
a modulated phase if the following criterion is satisfied, \cite{Axe70,Tagantsev13}
\begin{equation}
G^* = G-\frac{f^2}{C} <0, \label{Eq:CriterionFlexo}
\end{equation}
where $G^*$ is the renormalized correlation coefficient
after imposing the stationary condition on the strain \cite{Yudin13}.
Whenever Eq.~(\ref{Eq:CriterionFlexo}) holds, domain walls can spontaneously form, making the homogeneous ferroelectric phase unstable;
thus, $G^*$ can be regarded as a useful indicator of the proximity to the unstable regime.
To check whether Eq.~(\ref{Eq:CriterionFlexo}) is satisfied in PZO, we extract the relevant modulation-direction-dependent material parameters from first principles via a perturbative expansion of the energy around the cubic phase with respect to \emph{both} phonon mode amplitudes and wave vector $\hat{\bf q}$ \cite{Royo19} (see Supplemental Material 1 \cite{SM} for details) and obtain the full anisotropic $G^*(\hat{\bf q})$ dependence.
We find the absolute minimum of $G^*(\hat{\bf q})$ at $\hat{\bf q}=\langle110\rangle$, 82\% smaller than its maximum at $\hat{\bf q}=\langle100\rangle$, which matches well the modulation direction of the observed $Pbam$ structure.
However, though $G^*$ is significantly smaller than in other well-known perovskites such as BaTiO$_3$ and SrTiO$_3$ \cite{SM}, it is still positive -- meaning that a modulated state cannot form spontaneously in PZO due to flexoelectricity alone.

{\em Rotopolar coupling.}
There are other Lifshitz-like invariants that are allowed by symmetry in perovskites \cite{Eliseev13}:
for example the ``rotopolar'' coupling \cite{Schiaffino17}, whereby gradients of the AFD tilt modes,
rather than of the elastic strain, couple to the polarization.
In the context of the $Pbam$ phase of PZO, such coupling can be written as
\begin{equation}
  E_{\rm rp}=-WP_y\phi_y\frac{\partial\phi_x}{\partial x}.\label{rp_1d}
\end{equation}
This means that, in presence of a uniform transverse tilt ($\phi_y$), the gradient of the
transverse polarization ($P_y$) couples to the longitudinal tilt ($\phi_x$) and vice versa.
This fits nicely with the physical picture that emerges from Fig.~\ref{Fig:AFE-Decompose}(b):
both the observed 90$^\circ$ phase shift between the two modulated modes (a feature of an IC
phase transition driven by \emph{gradient} couplings \cite{Pottker16}) and the Cartesian
components of the distortions are consistent with Eq.~(\ref{rp_1d}).
For a quantitative assessment, we extract the numerical value of $W$ from first principles by
following the same procedure as in Ref.~\citenum{Schiaffino17}, i.e., by taking the long-wavelength
limit of the third-order force constants within the cubic reference structure.
Similar to $G^*$, the stationary condition on $\varepsilon_6=\partial u_y/\partial x$ leads to the renormalization of the rotopolar coefficient, $W^*=W+fR/C$, where $R$ quantifies the rotostrictive coupling, $E_{\rm rs}=-R\phi_x\phi_y \varepsilon_6$ -- see Supplemental Material 4 \cite{SM} for details.
We find that $W^*$ in PZO is about as large as in SrTiO$_3$, its anisotropy analysis gives the absolute maximum of $W^*(\hat{\bf q})$ along $\langle110\rangle$, 2.3 times stronger than $W^*(\langle100\rangle)$ \cite{SM}, consistent with the geometry of the AFE ground state.

As can be seen from Eq.~\eqref{rp_1d}, $E_{\rm rp}$ is manifestly a \emph{trilinear} function of the mode amplitudes,
which suggests some close connection to the model of Ref.~\citenum{Iniguez14}.
Indeed, as we identified above, $\Sigma$-, $R$- and $S$-point distortions are largely comprised of, respectively,
$P_y=Q_\Sigma\cos(qx)$, $\phi_y=Q_R$ and $\phi_x=Q_S\sin(qx)$, where $q=\pi/(\sqrt2a_0)$ is the modulation wavevector, $a_0$ is the lattice constant of the cubic unit cell.
Using this in Eq.~\eqref{rp_1d} we obtain $\langle E_{\rm rp}\rangle=-(W^*q/2)Q_\Sigma Q_RQ_S$, thus
recovering the trilinear $\Sigma$--$R$--$S$ coupling described in Ref.~\citenum{Iniguez14}.
After converting to the unit conventions of Ref.~\citenum{Iniguez14}, we obtain a
coupling energy between $\Sigma$, $R$ and $S$ modes amounting to 63.3~meV per 5-atom
formula unit (f.u.); this compares well with their reported value of 48.4~meV/f.u.

\begin{table}
  \centering
  \begin{tabular}{cr|cc}
  \multicolumn2{l|}{Structure}&DFT&Landau\\ \hline
  $Imcm$&uniform $\bm\phi||\langle110\rangle$&$-226$&$-226$\\
  $Ima2$&uniform $\mathbf P,\bm\phi||\langle110\rangle$&$-257$&$-259$\\
  $R3c$&uniform $\mathbf P,\bm\phi||\langle111\rangle$&$-275$&$-275$\\
  FiE&modulated $\mathbf P,\bm\phi||\langle110\rangle$&$-273$&$-265$\\
  $Pbam$&modulated $\mathbf P,\bm\phi||\langle110\rangle$&$-269$&$-273$\\
  4$\uparrow$4$\downarrow$&modulated $\mathbf P,\bm\phi||\langle110\rangle$&$-263$
    &$-268$
  \end{tabular}
  \caption{Energies of various phases in PZO compared to the cubic $Pm\bar3m$ structure in meV/f.u., with lattice parameters constrained to the $Pm\bar3m$ phase values.}\label{Tab:Energies}
\end{table}

{\em Continuum functional.}
To make the above arguments more quantitative, we construct an LGD functional
that incorporates the main physical ingredients discussed so far \cite{SM},
\begin{equation}
\begin{gathered}
  F=F_\text{hom}^*(\mathbf{P},\bm\phi)+\frac12G^*\Big(\frac{\partial P_y}{\partial x}\Big)^2\\
  +\frac12D_{11}\Big(\frac{\partial\phi_x}{\partial x}\Big)^2
    +\frac12D_{66}\Big(\frac{\partial\phi_y}{\partial x}\Big)^2
    -W^*P_y\phi_y\frac{\partial\phi_x}{\partial x}.
\end{gathered}\label{Eq:Landau}
\end{equation}
Here, $F_\text{hom}^*(\mathbf{P},\bm\phi)$ is the 8th-order expansion with respect to homogeneous $\bf P$ and $\bm\phi$,
obtained via a least-square fitting to the first-principles energy landscape of PZO at the $Pm\bar3m$ lattice parameters (the fixed-strain approach is justified for the purposes of this Letter, which will be shown below);
its description of the metastable non-modulated phases observed {\em ab initio} is essentially exact ($<3$~meV/f.u., see Tab.~\ref{Tab:Energies} and Supplemental Material 3 \cite{SM}).
The remaining terms in Eq.~\eqref{Eq:Landau} are specific to the $Pbam$ and other modulated phases of PZO, defined by the correlation coefficients for the polarization ($G^*$), for the longitudinal ($D_{11}$) and the transverse ($D_{66}$) tilts, and by the rotopolar coupling ($W^*$).

\begin{figure}
  \centering
  \includegraphics[width=\linewidth]{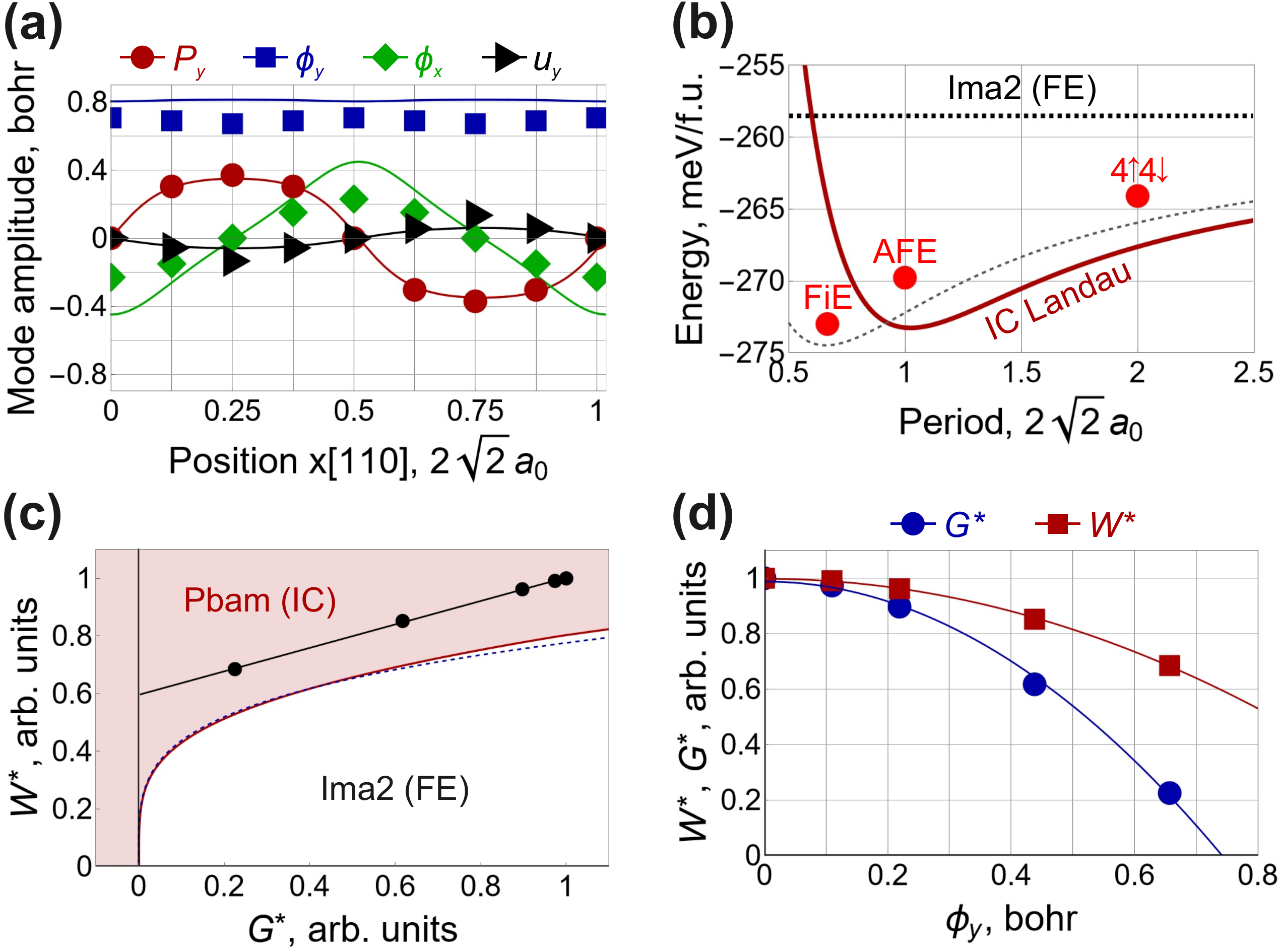}
  \caption{(a) Spatial distribution of polarization $P_y$, AFD tilts $\phi_x$, $\phi_y$, and displacement field $u_y$ in the IC $Pbam$ phase described by Eq.~\eqref{Eq:Landau} [solid lines], compared to the layer-by-layer decomposition shown in Fig.~\ref{Fig:AFE-Decompose}(b) [symbols]. 1~bohr of amplitude corresponds to 174~$\mu\rm C/\rm{cm}^2$ of polarization and $14.8^\circ$ of octahedral tilt. (b) Energy of the IC $Pbam$ phase at different periods of modulations described by Eq.~\eqref{Eq:Landau} with gradient coefficients extracted in $Pm\bar3m$ phase [red solid line] and in presence of a $\phi_y=0.7$~bohr distortion [gray dashed line], compared to the energies of the FE $Ima2$ phase [black dotted line] and of the modulated phases observed {\em ab initio} [red circles]. (c) Phase diagram between the $Pbam$ (red region) and $Ima2$ (white region) phases described by Eq.~\eqref{Eq:Landau} at varying correlation $G^*$ and rotopolar $W^*$ coefficients. Blue dotted line follows the analytical expression in Eq.~\eqref{Eq:RotoFlexoCond}. The black symbols and line correspond to the ones shown in (d). (d) $G^*$ and $W^*$ coefficients extracted from DFT at varying frozen uniform tilt $\phi_y$ (symbols), and their quadratic fits (solid lines).}\label{Fig:IC-Continuous}
\end{figure}

Minimization of Eq.~\eqref{Eq:Landau} via numerical finite-element methods yields a locally stable
modulated phase whose energy ($-273$~meV/f.u.) and equilibrium structure [see Fig.~\ref{Fig:IC-Continuous}(a)] accurately reproduce our DFT simulations of the AFE $Pbam$ phase.
Moreover, the optimal period of modulations following from our Landau potential lies within 5\% from the
experimentally observed $2\sqrt2a_0$ [Fig.~\ref{Fig:IC-Continuous}(a,b)], which supports the
{\em Umklapp} lock-in mechanism \cite{Bak82} that was proposed in earlier works \cite{Tagantsev13}.
The overall agreement between the DFT and continuum description of the $Pbam$ phase is exceptional,
given the fact that the modulations occur on the scale of few unit cells, i.e., in a regime where
the continuum approximation is usually regarded as unreliable.
We emphasise that such a close correspondence was obtained by constructing the Landau potential
of Eq.~\eqref{Eq:Landau} via a series of well-defined approximations to the reference DFT model
(i.e., by taking a rigorous long-wave limit of the low-energy Hamiltonian) and without introducing
any phenomenological fitting parameter to the gradient terms.

By its nature rotopolar coupling is not limited to the $Pbam$ phase but defines the whole range of modulated structures.
To show this we have performed additional {\em ab initio} and Landau-based analysis of the $\uparrow\uparrow\downarrow$ ferrielectric (FiE) structure, which has recently been described theoretically \cite{Aramberri21} and observed experimentally \cite{Jiang23}, and of the ``4-up-4-down'' modulated phase having modulation period twice as large as in $Pbam$ (Tab.~\ref{Tab:Energies}).
For the FiE phase, the first principles calculations show strong structural similarities with the $\uparrow\uparrow\downarrow\downarrow$ $Pbam$ phase, including uniform transverse and modulated longitudinal AFD tilts \cite{SM}, allowing its treatment as a modulated phase.
Our continuum theory matches well the {\em ab initio} energies of the three modulated phases (Tab.~\ref{Tab:Energies}), though the energy vs. period curve [red solid line in Fig.~\ref{Fig:IC-Continuous}(b)] is shifted towards larger modulation periods compared to the first-principles data points.
Note that our description of modulated phases is not merely theoretical: the available X-ray data on the {\em real} IC structures (e.g., in PbHfO$_3$ \cite{Fujishita18}, a closely related perovskite) show all features discussed in our Letter, including the modulated longitudinal tilts -- see Supplementary Material 7 \cite{SM} for details.

{\em Stability analysis.}
To formalise the role of the gradient couplings in the formation of modulated structures we study the stability of the ferroelectric (FE) $Ima2$ phase against an IC transition as a function of main physical parameters at play, $W^*$ and $G^*$.
$Ima2$, characterised by $\langle110\rangle$-oriented uniform polarization and AFD tilts, is essentially the modulationless ``parent phase'' of $Pbam$; therefore, the $Pbam$--$Ima2$ phase diagram will provide the \emph{necessary} condition for the IC phase to exist.
Note that strain relaxation (which is not included into our model potential) plays virtually no role in this context: our {\em ab initio} calculations show that FE $Ima2$, AFE $Pbam$ and FiE phases gain the same amount of energy at elastic equilibrium compared to the $Pm\bar3m$ lattice parameters, $-6$ to $-7$ meV/f.u.
Figure~\ref{Fig:IC-Continuous}(c) shows the $Pbam$--$Ima2$ phase boundary (red solid line) emerging from our direct study of the $Pbam$ phase energetics via minimisation of Eq.~\eqref{Eq:Landau} with respect to the order parameters and periodicity.

In the small-$G^*$ regime, the condition for the IC instability of $Ima2$ is well described [blue dotted line in Fig.~\ref{Fig:IC-Continuous}(c)]
by
\begin{equation}
  \gamma \left(\phi_{y0}W^*\right)^4>D_{11}G^*, \label{Eq:RotoFlexoCond}
\end{equation}
where $\phi_{y0}$ is the spontaneous tilt of the $Ima2$ phase, and $\gamma\sim1/[(\partial^2F_{\rm hom}^*/\partial P_y^2)(\partial^2F_{\rm hom}^*/\partial\phi_x^2)]$ is a well-defined ``softness'' parameter (see Supplemental Material 6 \cite{SM} for its exact expression).
Equation~\eqref{Eq:RotoFlexoCond} generalizes the classic criterion for the IC phase formation shown in Eq.~\eqref{Eq:CriterionFlexo},
recovering it for vanishing $\phi_{y0}$ or $W^*$.
The calculated $G^*$ and $W^*$ coefficients clearly satisfy Eq.~\eqref{Eq:RotoFlexoCond} in PZO
[$(1,1)$ point in Fig.~\ref{Fig:IC-Continuous}(c)], consistent with
the conclusions that we have reached via numerical minimization of Eq.~(\ref{Eq:Landau}).

A potential limitation of our approach consists in having calculated $G^*$ and $W^*$ in the $Pm\bar3m$ reference structure.
This raises the obvious question of whether the (large) uniform O$_6$ tilts may modify the value of such coefficients, possibly affecting the stability regime.
To verify such a possibility, we have recalculated both coefficients several times
in presence of a $\phi_y$ distortion of increasing amplitude, ranging from
zero to the ground-state value of Fig.~\ref{Fig:AFE-Decompose}(b).
The results, plotted in Fig.~\ref{Fig:IC-Continuous}(d), show a significant decrease of both coefficients for increasing $\phi_y$.
Remarkably, after plotting $G^*(\phi_y)$ and $W^*(\phi_y)$ in Fig.~\ref{Fig:IC-Continuous}(c),
the resulting line lies well within the unstable region of the phase diagram,
meaning that the $\phi_y$-dependence of these coefficients, not taken into account in
our Landau potential, is largely irrelevant for the tendency towards a modulated state.
At the same time, using the coefficients extracted at $\phi_y=0.7$~bohr in our potential [Eq.~\eqref{Eq:Landau}] gives a closer match of the optimal periodicity of the IC phase [gray dashed line in Fig.~\ref{Fig:IC-Continuous}(b)].

It is interesting to note that $G^*$ transitions to negative values for tilt amplitudes
that are only slightly larger than our calculated $\phi_{y0}$.
Such a critical behavior suggests that the antiferroelectric state of PZO may, in fact,
originate from a \emph{triggered} IC transition [governed, at leading order in $\phi_y$,
by $E_{\rm tr} = K \phi_y^2 (\partial P_y/\partial x)^2$], thus questioning the
necessity of the rotopolar mechanism that we propose here.
(Note that this same coupling term, $E_{\rm tr}$, was proposed recently as a driving
force towards incommensuration in closely related materials \cite{Kniazeva22}.)
To confirm the role of the rotopolar mechanism in the $Pbam$ phase formation we have performed constrained-DFT relaxations where the secondary tilt $\phi_x$ (and hence the rotopolar coupling) is suppressed by hand.
This can be done cleanly as the modulated $\phi_x$ distortion has a different symmetry
compared to the other active modes in the system.
The energy of the resulting AFE $Pbam$-like structure is $-256$~meV/f.u., i.e., slightly
higher than $Ima2$, meaning that there is no driving force towards incommensuration in
the absence of the rotopolar coupling, consistent with Fig.~\ref{Fig:IC-Continuous}(c).
This result conclusively demonstrates the \emph{necessity} of the rotopolar coupling for
stabilizing the AFE state in PZO, making it the prime candidate for the ``missing''
mechanism alluded to in Ref.~\citenum{Tagantsev13}.

{\em Discussion and outlook.}
We further speculate that either the rotopolar mechanism or such higher-order couplings may be involved in the formation of the recently discussed $Pnma$--80 phase \cite{Baker21Arxiv,Grosso21}
(the additional distortions take indeed the form of 2D-modulated polarization and tilts), and thereby provide a unified framework to explain the emergence of spatially modulated structure in tilted perovskites.
Application of these ideas to the many examples that have been recently discussed in the literature \cite{Kim13,Bosak20,Fu20,Grosso21} constitutes, in our view, an exciting avenue for further study.

\begin{acknowledgments}
 We acknowledge the support of Ministerio de Economia,
 Industria y Competitividad (MINECO-Spain) through
 Grants  No.  MAT2016-77100-C2-2-P, No. PID2019-108573GB-C22
 and Severo Ochoa FUNFUTURE center of excellence (CEX2019-000917-S);
 and of Generalitat de Catalunya (Grant No. 2017 SGR1506).
 This project has received funding from the European
 Research Council (ERC) under the European Union's
 Horizon 2020 research and innovation program (Grant
 Agreement No. 724529). Part of the calculations were performed at
 the Supercomputing Center of Galicia (CESGA).
 We thank Philippe Ghosez for useful discussions and Roman Burkovsky for reading through the manuscript and providing input to the text.
\end{acknowledgments}

\bibliography{PZO.Incommensurate}



\section*{Supplementary material}

<div align="center">

**Supplemental Materials:**

**Tilt-driven antiferroelectricity in PbZrO₃**

</div>

K. Shapovalov[1] and M. Stengel[2, 1]

[1]*Institut de Ciència de Materials de Barcelona (ICMAB-CSIC),*

*Campus UAB, 08193 Bellaterra, Spain*

[2]*ICREA — Institució Catalana de Recerca i Estudis Avançats, 08010 Barcelona, Spain*

# Contents

# Supplemental Material 1. Details of first-principles calculations

Our calculations are performed in the framework of the local-density approximation to density functional theory as implemented in the ABINIT package [1–3]. The pseudopotentials are generated with the FHI package in the following electronic configurations:

Pb($5d^{10}6s^26p^0$), Zr($5s^25p^04d^2$), O($2s^22p^4$). The sampling of the Brillouin zone is performed with Monkhorst-Pack meshes whose specifics vary depending on the geometry of the cell; in all cases the density is equivalent or better to an $8 \times 8 \times 8$ sampling of the 5-atom primitive cell in the reference $Pm\bar{3}m$ phase. With the above parameters, we relax the atomic structures (and, in selected cases, the cell parameters) to an accuracy of 0.1 meV/f.u. or better. The resulting equilibrium lattice parameter of the cubic phase, which we use as reference in our calculations, is 4.098 Å.

The calculation of the gradient-mediated couplings, and in particular of the $f$, $G$ and $W$ coefficients, is performed on a 20-atom cell, within a reference frame that is rotated by $45°$ in the $xy$ plane with respect to the pseudocubic axes. In particular, we assume that the $x$, $y$ and $z$ axes are oriented along [110], [1$\bar{1}$0] and [001] directions. Therefore, the $x$ axis coincides with the modulation direction, while the polarization (Pb displacements) and the main tilt axis ($R$-mode) are oriented along $y$.

Starting from the atoms in their reference cubic structure, we generate a series of distorted geometries by freezing in by hand an increasing amplitude of the main tilt, $\phi_y$. (The values of $\phi_y$ range from zero to the spontaneous distortion of the ground-state $Pbam$ structure, both inclusive.) For each structure, we compute the force-constant matrices along a one-dimensional mesh of $\mathbf{q}$-points along the $x$ direction, via the linear-response module of ABINIT. (In practice we use the points 0, 1/6, 1/3 and 1/2, corresponding to a linear mesh of six $\mathbf{q}$-points.) The result is then Fourier-transformed back to real space to obtain the interatomic force constants (IFCs) along the modulation direction,

$$\Phi^l_{\kappa\alpha,\kappa'\beta} = \frac{\partial^2 E}{\partial R_{0\kappa\alpha}\partial R_{l\kappa'\beta}}. \tag{S1}$$

The latter is defined as second derivative of the total energy $E$ with respect to displacements of the atom $0\kappa$ along $\alpha$ and of the atom $l\kappa'$ along $\beta$. ($\kappa, \kappa'$ are sublattice indices, $l$ is a cell index.)

Next, we calculate the real-space moments of the IFCs via

$$\Phi^{(n)}_{\kappa\alpha,\kappa'\beta} = \sum_l \Phi^l_{\kappa\alpha,\kappa'\beta}(R^{(0)}_{l\kappa'x} - R^{(0)}_{0\kappa x})^n, \tag{S2}$$

where $\mathbf{R}^{(0)}_{l\kappa'}$ stands for the unperturbed position of the atom $l\kappa'$ within the cubic reference phase. (This means that $\mathbf{R}^{(0)}_{l\kappa'}$ does not contain the contribution of the $\phi_y$ tilts.) [$\Phi^{(0)}_{\kappa\alpha,\kappa'\beta}$

corresponds to the zone-center force-constants matrix of the distorted 20-atom cell by construction.] The $n = 2$ moments of the IFCs are used to calculate the $C$, $f$, $G$ coefficients following the prescriptions of Ref. 4,

$$C = -\frac{1}{2\Omega} \sum_{\kappa\kappa'} \Phi^{(2)}_{\kappa y, \kappa' y}, \tag{S3a}$$

$$f = -\frac{1}{2\Omega} \sum_{\kappa\kappa'} \Phi^{(2)}_{\kappa y, \kappa' \beta} p_{\kappa' \beta}, \tag{S3b}$$

$$G = -\frac{1}{2\Omega} \sum_{\kappa\kappa'} \Phi^{(2)}_{\kappa\alpha, \kappa' \beta} p_{\kappa\alpha} p_{\kappa' \beta}, \tag{S3c}$$

where $\Omega$ is the volume of the cell, and $p_{\kappa\alpha}$ is an appropriately normalized vector describing the uniform ferroelectric distortion along $y$. (The shear elastic constant $C$ can be also calculated by simpler means, and this provides an excellent test for the numerical quality of the lattice sums.)

Finally, the trilinear $W$ coefficient is extracted from the $n = 1$ term in the expansion following Ref. 5,

$$W \simeq \frac{1}{\phi_y \Omega} \sum_{\kappa\kappa'} \Phi^{(1)}_{\kappa\alpha, \kappa' \beta} \phi^{(x)}_{\kappa\alpha} p_{\kappa' \beta}. \tag{S4}$$

Here $\phi^{(x)}_{\kappa\alpha}$ denotes the (normalized) atomic displacement pattern that is associated with a uniform longitudinal tilt, and the $1/\phi_y$ prefactor takes care of the numerical differentiation (hence the "$\simeq$" sign) with respect to the main transverse tilt. We have also recalculated the $W$ coefficient by using the recent implementation of the long-wavelength module in ABINIT together with ONCVPSP pseudopotentials [6] finding excellent numerical agreement.

## Supplemental Material 2.   Real-space mapping between atomic distortions and continuum fields

**Direct mapping.** In order to extract the local inhomogeneous order parameters from the atomic distortions in the *Pbam* structure of PZO, we perform a layer-by-layer decomposition as follows. Focusing on (110) planes, there are two types of atomic layers in the 40-atoms unit cell: the ones comprised of 2Pb, 2Zr, 2O atoms and the ones comprised of 4O atoms (see Fig. S1). We denote the local atomic displacements within a layer $n$ as $\mathbf{v_n}$, which is a vector with $6 \times 3 = 18$ components for 2Pb, 2Zr, 2O layers and $4 \times 3 = 12$ components

for 4O layers. Our goal is to decompose these displacements over phonon modes which we index with $\alpha$: polar, acoustic, antiferrodistortive (AFD) modes.

Let atomic displacements produced by a mode $\alpha$ within layer $n$ be $\mathbf{e_{n\alpha}}$ (also a vector with either 12 or 18 components). The phonon modes are orthogonal over any two neighbouring layers:

$$(\mathbf{e_{n\alpha}}, \mathbf{e_{n\beta}}) + (\mathbf{e_{n+1\,\alpha}}, \mathbf{e_{n+1\,\beta}}) = \epsilon_\alpha^2 \delta_{\alpha\beta}, \tag{S5}$$

where $(\cdot, \cdot)$ is the scalar product calculated within one layer, $\epsilon_\alpha$ is the norm of the mode calculated over two layers (independent of $n$ by the choice of the phonon modes), and $\delta_{\alpha\beta}$ is the Kronecker delta. Then one can decompose the atomic displacements $\mathbf{v_n}$ over basis $\mathbf{e_{n\alpha}}$ as follows:

$$\mathbf{v_n} = \sum_\alpha k_{n\alpha} \mathbf{e_{n\alpha}} + \mathbf{r_n}, \tag{S6}$$

where $\mathbf{r_n}$ is the remainder. Minimization of the remainder leads to the following coefficients $k_{n\alpha}$:

$$k_{n\alpha} = \sum_\beta (E_{n\alpha\beta})^{-1} V_{n\beta}, \tag{S7}$$

$$E_{n\alpha\beta} = (\mathbf{e_{n\alpha}}, \mathbf{e_{n\beta}}), \qquad V_{n\beta} = (\mathbf{v_n}, \mathbf{e_{n\beta}}), \tag{S8}$$

where it is assumed that all studied phonon modes are characterized by non-zero displacement $\mathbf{e_{n\alpha}}$ in each layer.

Observable order parameters, such as polarization, AFD tilts etc., are defined by displacements of all five atoms of the $PbZrO_3$ formula unit. Since coefficients $k_{n\alpha}$ are defined by displacements of only 60% of the formula unit (2Pb, 2Zr, 2O layers) or 40% of the f.u. (4O layers), they cannot be considered as actual representations of the observable order parameters. In order to take into account the whole formula unit, we modify Eqs. (S7,S8) by projecting atomic displacements onto a basis spanning over three atomic layers multiplied by the lattice-independent $(1/2, 1, 1/2)$ envelope:

$$p_{n\alpha} = \sum_\beta (\hat{E}_{n\alpha\beta})^{-1} \hat{V}_{n\beta}, \tag{S9}$$

$$\hat{E}_{n\alpha\beta} = \langle \mathbf{e_{n\alpha}}, \mathbf{e_{n\beta}} \rangle, \qquad \hat{V}_{n\beta} = \langle \mathbf{v_n}, \mathbf{e_{n\beta}} \rangle, \tag{S10}$$

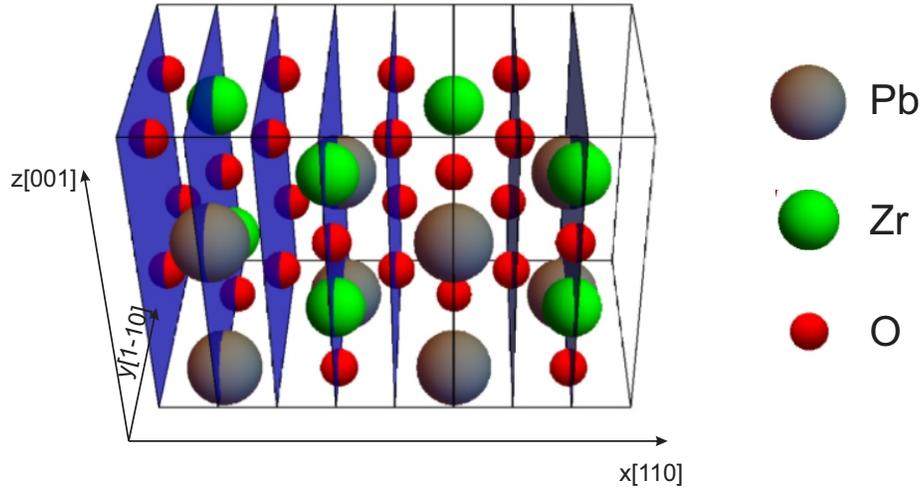

FIG. S1. Atomic layers of PZO used in layer-by-layer composition, composed alternately of 2Pb, 2Zr, 2O atoms and of 4O atoms.

where $\langle \cdot, \cdot \rangle$ is the "extended" scalar product taking into account displacements of all five atoms within the formula unit of PbZrO$_3$, defined as

$$\langle \mathbf{v_n}, \mathbf{w_n} \rangle = \frac{1}{2}(\mathbf{v_{n-1}}, \mathbf{w_{n-1}}) + (\mathbf{v_n}, \mathbf{w_n}) + \frac{1}{2}(\mathbf{v_{n+1}}, \mathbf{w_{n+1}}). \tag{S11}$$

By the choice of the phonon modes, $(\mathbf{e_{n\alpha}}, \mathbf{e_{n\alpha}}) = (\mathbf{e_{n+2\,\alpha}}, \mathbf{e_{n+2\,\alpha}})$. As a consequence,

$$\hat{E}_{n\alpha\beta} = \frac{1}{2}(\mathbf{e_{n-1\,\alpha}}, \mathbf{e_{n-1\,\beta}}) + (\mathbf{e_{n\alpha}}, \mathbf{e_{n\beta}}) + \frac{1}{2}(\mathbf{e_{n+1\,\alpha}}, \mathbf{e_{n+1\,\beta}}) = \epsilon_\alpha^2 \delta_{\alpha\beta}, \tag{S12}$$

and Eq. (S9) becomes

$$p_{n\alpha} = \frac{1}{\epsilon_\alpha^2} \left( \frac{1}{2} V_{n-1\,\alpha} + V_{n\alpha} + \frac{1}{2} V_{n+1\,\alpha} \right), \tag{S13}$$

which concludes the direct mapping between atomic distortions and order parameters.

**Converse mapping.** To extract atomic distortions $\mathbf{v_n}$ from local order parameter am-

plitudes $p_{n\alpha}$, one needs first to solve the matrix equation following from Eq. (S13):

$$\mathbf{A}\mathbf{V_\alpha} = \epsilon_\alpha^2 \mathbf{p_\alpha}, \tag{S14}$$

$$\mathbf{A} = \begin{pmatrix} 1 & \frac{1}{2} & 0 & 0 & 0 & 0 & 0 & \frac{1}{2} \\ \frac{1}{2} & 1 & \frac{1}{2} & 0 & 0 & 0 & 0 & 0 \\ 0 & \frac{1}{2} & 1 & \frac{1}{2} & 0 & 0 & 0 & 0 \\ 0 & 0 & \frac{1}{2} & 1 & \frac{1}{2} & 0 & 0 & 0 \\ 0 & 0 & 0 & \frac{1}{2} & 1 & \frac{1}{2} & 0 & 0 \\ 0 & 0 & 0 & 0 & \frac{1}{2} & 1 & \frac{1}{2} & 0 \\ 0 & 0 & 0 & 0 & 0 & \frac{1}{2} & 1 & \frac{1}{2} \\ \frac{1}{2} & 0 & 0 & 0 & 0 & 0 & \frac{1}{2} & 1 \end{pmatrix}, \tag{S15}$$

$$\mathbf{V_\alpha} = (V_{1\alpha}, V_{2\alpha}, ..., V_{8\alpha}), \qquad \mathbf{p_\alpha} = (p_{1\alpha}, p_{2\alpha}, ..., p_{8\alpha}). \tag{S16}$$

In the atomic structure of the *Pbam* phase, there are eight atomic layers periodically repeated, and $\mathbf{A}$ is therefore an $8 \times 8$ matrix.

Strictly speaking, the inverse of $\mathbf{A}$ is undefined since $\det \mathbf{A} = 0$, meaning that $\mathbf{v_n} \leftrightarrow p_{n\alpha}$ is not an isomorphism. To determine which information is lost during the direct mapping, we study the eigenvalues and eigenvectors of $\mathbf{A}$: it has seven non-zero eigenvalues, and the only zero eigenvalue corresponds to the eigenvector $(1, -1, 1, -1, 1, -1, 1, -1)$. Taking this into account, one can calculate the pseudoinverse $\mathbf{A}^+$ using well-established methods, and obtain $\mathbf{V_\alpha}$:

$$\mathbf{V_\alpha} = \epsilon_\alpha^2 \mathbf{A}^+ \mathbf{p_\alpha} + C(1, -1, 1, -1, 1, -1, 1, -1), \tag{S17}$$

where we assume $C = 0$ for our converse mapping. The atomic distortions $\mathbf{v_n}$ can then be obtained from $\mathbf{V_\alpha}$ via Eqs. (S6,S7).

## Supplemental Material 3. Landau energy: homogeneous phases

For our Landau-based calculations, we use the following 8th-order energy expansion for the homogeneous energy $F_{\text{hom}}(\mathbf{P}, \boldsymbol{\phi})$, at varying polar mode amplitude $\mathbf{P} = (P_1, P_2, P_3)$ and AFD mode amplitude $\boldsymbol{\phi} = (\phi_1, \phi_2, \phi_3)$, incorporating symmetry features of the prototypical $Pm\bar{3}m$ phase:

$$F_{\text{hom}}(\mathbf{P}, \boldsymbol{\phi}) = F_P(\mathbf{P}) + F_\phi(\boldsymbol{\phi}) + F_{P\phi}(\mathbf{P}, \boldsymbol{\phi}), \tag{S18}$$

$$F_P(\mathbf{P}) = \frac{\alpha_1}{2}(P_1^2 + P_2^2 + P_3^2) + \frac{\alpha_{11}}{4}(P_1^4 + P_2^4 + P_3^4) + \frac{\alpha_{12}}{2}(P_1^2 P_2^2 + P_2^2 P_3^2 + P_3^2 P_1^2)$$

$$+ \frac{\alpha_{111}}{6}(P_1^6 + P_2^6 + P_3^6) + \alpha_{112}\Big(P_1^4(P_2^2 + P_3^2) + P_2^4(P_3^2 + P_1^2) + P_3^4(P_1^2 + P_2^2)\Big)$$

$$+ \alpha_{123}P_1^2 P_2^2 P_3^2 + \frac{\alpha_{1111}}{8}(P_1^8 + P_2^8 + P_3^8)$$

$$+ \alpha_{1112}\Big(P_1^6(P_2^2 + P_3^2) + P_2^6(P_3^2 + P_1^2) + P_3^6(P_1^2 + P_2^2)\Big)$$

$$+ \alpha_{1122}(P_1^4 P_2^4 + P_2^4 P_3^4 + P_3^4 P_1^4) + \alpha_{1123}(P_1^4 P_2^2 P_3^2 + P_2^4 P_3^2 P_1^2 + P_3^4 P_1^2 P_2^2), \tag{S19}$$

$$F_\phi(\boldsymbol{\phi}) = \frac{\beta_1}{2}(\phi_1^2 + \phi_2^2 + \phi_3^2) + \frac{\beta_{11}}{4}(\phi_1^4 + \phi_2^4 + \phi_3^4) + \frac{\beta_{12}}{2}(\phi_1^2 \phi_2^2 + \phi_2^2 \phi_3^2 + \phi_3^2 \phi_1^2)$$

$$+ \frac{\beta_{111}}{6}(\phi_1^6 + \phi_2^6 + \phi_3^6) + \beta_{112}\Big(\phi_1^4(\phi_2^2 + \phi_3^2) + \phi_2^4(\phi_3^2 + \phi_1^2) + \phi_3^4(\phi_1^2 + \phi_2^2)\Big)$$

$$+ \beta_{123}\phi_1^2 \phi_2^2 \phi_3^2 + \frac{\beta_{1111}}{8}(\phi_1^8 + \phi_2^8 + \phi_3^8)$$

$$+ \beta_{1112}\Big(\phi_1^6(\phi_2^2 + \phi_3^2) + \phi_2^6(\phi_3^2 + \phi_1^2) + \phi_3^6(\phi_1^2 + \phi_2^2)\Big)$$

$$+ \beta_{1122}(\phi_1^4 \phi_2^4 + \phi_2^4 \phi_3^4 + \phi_3^4 \phi_1^4) + \beta_{1123}(\phi_1^4 \phi_2^2 \phi_3^2 + \phi_2^4 \phi_3^2 \phi_1^2 + \phi_3^4 \phi_1^2 \phi_2^2), \tag{S20}$$

$$F_{P\phi}(\mathbf{P}, \boldsymbol{\phi}) = Q_{11}(P_1^2 \phi_1^2 + P_2^2 \phi_2^2 + P_3^2 \phi_3^2) + Q_{12}\Big(P_1^2(\phi_2^2 + \phi_3^2)$$

$$+ P_2^2(\phi_3^2 + \phi_1^2) + P_3^2(\phi_1^2 + \phi_2^2)\Big) + Q_{44}(P_1 P_2 \phi_1 \phi_2 + P_2 P_3 \phi_2 \phi_3 + P_3 P_1 \phi_3 \phi_1)$$

$$+ S_{111}(P_1^4 \phi_1^2 + P_2^4 \phi_2^2 + P_3^4 \phi_3^2) + S_{112}\Big(P_1^4(\phi_2^2 + \phi_3^2) + P_2^4(\phi_3^2 + \phi_1^2) + P_3^4(\phi_1^2 + \phi_2^2)\Big)$$

$$+ S_{121}\Big(P_1^2 P_2^2(\phi_1^2 + \phi_2^2) + P_2^2 P_3^2(\phi_2^2 + \phi_3^2) + P_3^2 P_1^2(\phi_3^2 + \phi_1^2)\Big)$$

$$+ S_{123}(P_1^2 P_2^2 \phi_3^2 + P_2^2 P_3^2 \phi_1^2 + P_3^2 P_1^2 \phi_2^2)$$

$$+ S_{166}\Big((P_1^2 + P_2^2)P_1 P_2 \phi_1 \phi_2 + (P_2^2 + P_3^2)P_2 P_3 \phi_2 \phi_3 + (P_3^2 + P_1^2)P_3 P_1 \phi_3 \phi_1\Big)$$

$$+ S_{144}(P_1^2 P_2 P_3 \phi_2 \phi_3 + P_2^2 P_3 P_1 \phi_3 \phi_1 + P_3^2 P_1 P_2 \phi_1 \phi_2)$$

$$+ T_{111}(\phi_1^4 P_1^2 + \phi_2^4 P_2^2 + \phi_3^4 P_3^2) + T_{112}\Big(\phi_1^4(P_2^2 + P_3^2) + \phi_2^4(P_3^2 + P_1^2) + \phi_3^4(P_1^2 + P_2^2)\Big)$$

$$+ T_{121}\Big(\phi_1^2 \phi_2^2(P_1^2 + P_2^2) + \phi_2^2 \phi_3^2(P_2^2 + P_3^2) + \phi_3^2 \phi_1^2(P_3^2 + P_1^2)\Big)$$

$$+ T_{123}(\phi_1^2 \phi_2^2 P_3^2 + \phi_2^2 \phi_3^2 P_1^2 + \phi_3^2 \phi_1^2 P_2^2)$$

$$+ T_{166}\Big((\phi_1^2 + \phi_2^2)\phi_1 \phi_2 P_1 P_2 + (\phi_2^2 + \phi_3^2)\phi_2 \phi_3 P_2 P_3 + (\phi_3^2 + \phi_1^2)\phi_3 \phi_1 P_3 P_1\Big)$$

$$+ T_{144}(\phi_1^2 \phi_2 \phi_3 P_2 P_3 + \phi_2^2 \phi_3 \phi_1 P_3 P_1 + \phi_3^2 \phi_1 \phi_2 P_1 P_2). \tag{S21}$$

Here, indices 1, 2, 3 correspond to [100], [010], [001] pseudocubic crystallographic directions.

The numerical values of the coefficients in Eqs. (S18–S21) are listed in Table S-I where atomic units (Ha, bohr) are used and both polarization $\mathbf{P}$ and tilts $\boldsymbol{\phi}$ are measured in bohrs. These values were extracted by conducting a series of constrained DFT calculations of PZO with varying frozen polar distortions $\mathbf{P}$ and AFD distortions $\boldsymbol{\phi}$ and fitting Eqs. (S18–S21) to the obtained energy landscape. 1 bohr of amplitude of $\mathbf{P}$ in [100] direction corresponds to 1.721 bohr of Pb displacement, 0.343 bohr of Zr, $-0.111$ bohr of apical oxygen, and $-0.977$ bohr of equatorial oxygens displacements in [100] direction, which is the soft polar eigenvector of the force-constant matrix that we extracted from ab-initio. This choice of the polar mode is justified by the near-perfect description of the energy of metastable phases by the Landau potential, as we will show it below. 1 bohr of amplitude of $\boldsymbol{\phi}$ in [100] direction corresponds to 1 bohr of the displacement of the equatorial oxygens while the oxygen octahedra rotate around the [100] axis. The following series of constrained ab-initio calculations were conducted:

- $\left(|\mathbf{P}| = 0, 0.05, ..., 0.6 \text{ bohr}\right) \bigotimes \begin{pmatrix} \mathbf{P}||[100] \\ \mathbf{P}||[110] \\ \mathbf{P}||[111] \end{pmatrix}$ – three sets of calculations for $F_P(\mathbf{P})$;

- $\left(|\boldsymbol{\phi}| = 0, 0.1, ..., 1.2 \text{ bohr}\right) \bigotimes \begin{pmatrix} \boldsymbol{\phi}||[100] \\ \boldsymbol{\phi}||[110] \\ \boldsymbol{\phi}||[111] \end{pmatrix}$ – three sets of calculations for $F_\phi(\boldsymbol{\phi})$;

- $\begin{pmatrix} (|\mathbf{P}|, |\boldsymbol{\phi}|) = (0,0), (0.025, 0.1), ..., (0.3, 1.2) \\ (|\mathbf{P}|, |\boldsymbol{\phi}|) = (0,0), (0.05, 0.1), ..., (0.6, 1.2) \\ (|\mathbf{P}|, |\boldsymbol{\phi}|) = (0,0), (0.05, 0.05), ..., (0.6, 0.6) \end{pmatrix} \bigotimes \begin{pmatrix} \mathbf{P}||[100] \\ \mathbf{P}||[110] \end{pmatrix} \bigotimes \begin{pmatrix} \boldsymbol{\phi}||[100] \\ \boldsymbol{\phi}||[110] \end{pmatrix}$
  – $3 \times 2 \times 2 = 12$ sets of calculations for $F_{P\phi}(\mathbf{P}, \boldsymbol{\phi})$ taking into account interaction between $\mathbf{P}$ and $\boldsymbol{\phi}$ within (001) crystallographic plane;

- $\begin{pmatrix} (|\mathbf{P}|, |\boldsymbol{\phi}|) = (0,0), (0.025, 0.1), ..., (0.3, 1.2) \\ (|\mathbf{P}|, |\boldsymbol{\phi}|) = (0,0), (0.05, 0.1), ..., (0.6, 1.2) \\ (|\mathbf{P}|, |\boldsymbol{\phi}|) = (0,0), (0.05, 0.05), ..., (0.6, 0.6) \end{pmatrix} \bigotimes \left(\mathbf{P}||[111], \boldsymbol{\phi}||[111]\right)$
  – three sets of calculations for $F_{P\phi}(\mathbf{P}, \boldsymbol{\phi})$ taking into account interaction between $\mathbf{P}$ and $\boldsymbol{\phi}$ in [111] crystallographic direction.

The local energy minima in the resulting Landau energy (Eqs. S18–S21) describe with a high accuracy the energies of most phases that we calculated in DFT, with a notable

| | | | | |
|---|---|---|---|---|
| $\alpha_1$ | $-0.371 \times 10^{-3}$ | | $\beta_1$ | $-9.331 \times 10^{-5}$ |
| $\alpha_{11}$ | $3.029 \times 10^{-3}$ | | $\beta_{11}$ | $18.602 \times 10^{-5}$ |
| $\alpha_{12}$ | $1.451 \times 10^{-3}$ | | $\beta_{12}$ | $11.540 \times 10^{-5}$ |
| $\alpha_{111}$ | $-4.143 \times 10^{-3}$ | | $\beta_{111}$ | $-7.864 \times 10^{-5}$ |
| $\alpha_{112}$ | $1.010 \times 10^{-3}$ | | $\beta_{112}$ | $-2.626 \times 10^{-5}$ |
| $\alpha_{123}$ | $-3.700 \times 10^{-3}$ | | $\beta_{123}$ | $9.064 \times 10^{-5}$ |
| $\alpha_{1111}$ | $3.814 \times 10^{-3}$ | | $\beta_{1111}$ | $2.409 \times 10^{-5}$ |
| $2\alpha_{1112} + \alpha_{1122}$ | $4.994 \times 10^{-3}$ | | $2\beta_{1112} + \beta_{1122}$ | $1.278 \times 10^{-5}$ |
| $\alpha_{1123}$ | $-0.450 \times 10^{-3}$ | | $\beta_{1123}$ | $0.103 \times 10^{-5}$ |
| $Q_{11}$ | $-1.623 \times 10^{-4}$ | | | |
| $Q_{12}$ | $-3.124 \times 10^{-4}$ | | | |
| $Q_{44}$ | $2.268 \times 10^{-4}$ | | | |
| $S_{111}$ | $0.551 \times 10^{-3}$ | | $T_{111}$ | $0.302 \times 10^{-4}$ |
| $S_{112}$ | $-0.402 \times 10^{-3}$ | | $T_{112}$ | $-0.945 \times 10^{-4}$ |
| $S_{121}$ | $1.273 \times 10^{-3}$ | | $T_{121}$ | $1.794 \times 10^{-4}$ |
| $S_{166}$ | $-2.294 \times 10^{-3}$ | | $T_{166}$ | $-1.922 \times 10^{-4}$ |
| $S^{[111]}$ | $-0.149 \times 10^{-3}$ | | $T^{[111]}$ | $-0.563 \times 10^{-4}$ |

TABLE S-I. Tensor components in Eqs. (S18–S21) describing homogeneous phases of PZO; indices 1,2,3 correspond to [100], [010], [001] pseudocubic crystallographic directions. All values shown are in atomic units (Ha, bohr); both polarization $\mathbf{P}$ and tilts $\boldsymbol{\phi}$ are measured in bohrs. Designations $S^{[111]} = (S_{111} + 2S_{1112} + 2S_{121} + S_{123} + S_{144} + 2S_{166})/9$ and $T^{[111]} = (T_{111} + 2T_{1112} + 2T_{121} + T_{123} + T_{144} + 2T_{166})/9$ for the tensor components appearing in $R3c$ phase geometry are used.

exception of $Imcm$ ($||\phi[110]$) – see Table S-II. We relate this discrepancy to the additional R-mode Pb AFE distortions that appear in the relaxed $Imcm$ structure, coupled to the AFD distortions $\boldsymbol{\phi}$ [7]. It can be seen from the symmetry analysis that the additional energy associated to this AFE mode ($\boldsymbol{\eta}$) can be written as:

$$F_{\text{AFE}}(\eta_1, \eta_2, \eta_3; \phi_1, \phi_2, \phi_3) = \frac{\alpha_{\text{AFE}}}{2}(\eta_1^2 + \eta_2^2 + \eta_3^2)$$
$$+ k\left(\eta_1\phi_1(\phi_3^2 - \phi_2^2) + \eta_2\phi_2(\phi_1^2 - \phi_3^2) + \eta_3\phi_3(\phi_2^2 - \phi_1^2)\right) + K_{ijkl}\eta_i\eta_j\eta_k\phi_l + \ldots$$

(S22)

In order to estimate the effect of the AFE distortions on the energy we take the first

| Structure | | | Landau | | DFT | |
|-----------|---|---|---------|---|-----|---|
| | | | No AFE ($\boldsymbol{\eta} = 0$) | With AFE ($\boldsymbol{\eta} \neq 0$) | Constrained | Relaxed |
| $P4mm$ | $\mathbf{P}[100]$ | | $-162$ | $-162$ | $-162$ | |
| $Amm2$ | $\mathbf{P}[110]$ | | $-187$ | $-187$ | $-190$ | |
| $I4/mcm$ | | $\boldsymbol{\phi}[100]$ | $-174$ | $-174$ | $-174$ | |
| $Imcm$ | | $\boldsymbol{\phi}[110]$ | $-203$ | $-226$ | $-226$ | $-243$ |
| $R\bar{3}c$ | | $\boldsymbol{\phi}[111]$ | $-214$ | $-214$ | $-214$ | $-227$ |
| $Ima2$ | $\mathbf{P}[110]$ | $\boldsymbol{\phi}[110]$ | $-253$ | $-259$ | $-257$ | $-264$ |
| $R3c$ | $\mathbf{P}[111]$ | $\boldsymbol{\phi}[111]$ | $-275$ | $-275$ | $-275$ | $-275$ |

| | $\beta_{112}^{\dagger}$ $\beta_{123}^{\dagger}$ | $\beta_{112}$ $\beta_{123}$ | |
|---|---|---|---|
| | $-1.140$ $0.149$ | $-2.626$ $9.064$ | $\times 10^{-5}$ |

TABLE S-II. Energies of various phases in PZO with and without taking into account AFE mode $\boldsymbol{\eta}$ in meV per formula unit (f.u.). $\beta_{112}^{\dagger}$, $\beta_{123}^{\dagger}$ are the values extracted from ab-initio without taking into account AFE mode $\boldsymbol{\eta}$. $\beta_{112}$, $\beta_{123}$ are the values renormalized according to Eq. (S24).

two terms in Eq. (S22) and neglect the impact of $K_{ijkl}\eta_i\eta_j\eta_k\phi_l$ and higher-order terms. Minimizing the energy allows us to rewrite Eq. (S22) as

$$F_{\text{AFE}}(\phi_1, \phi_2, \phi_3) \approx \delta\beta\Big(\phi_1^4(\phi_2^2 + \phi_3^2) + \phi_2^4(\phi_3^2 + \phi_1^2) + \phi_3^4(\phi_1^2 + \phi_2^2) - 6\phi_1^2\phi_2^2\phi_3^2\Big),$$
$$\delta\beta = -\frac{k^2}{2\alpha_{\text{AFE}}},$$
$$\eta_1 \approx -\frac{k}{\alpha_{\text{AFE}}}\phi_1(\phi_3^2 - \phi_2^2), \quad \eta_2 \approx -\frac{k}{\alpha_{\text{AFE}}}\phi_2(\phi_1^2 - \phi_3^2), \quad \eta_3 \approx -\frac{k}{\alpha_{\text{AFE}}}\phi_3(\phi_2^2 - \phi_1^2).$$
$$\text{(S23)}$$

From Eq. (S23) one can see that [100] and [111] AFD modes $\boldsymbol{\phi}$ do not trigger R-mode AFE distortions, while in the case of [110] AFD mode the AFE mode $\boldsymbol{\eta}$ becomes active and has an impact on the overall energy. To take into account the effect of Eq. (S23) on the total energy of the system we renormalize $\beta_{112}$ and $\beta_{123}$ coefficients as follows:

$$\beta_{112} = \beta_{112}^{\dagger} + \delta\beta, \qquad \beta_{123} = \beta_{123}^{\dagger} - 6\delta\beta, \qquad \text{(S24)}$$

where $\beta_{112}^{\dagger}$ and $\beta_{123}^{\dagger}$ are the coefficients that we extracted previously without taking into account the R-point AFE mode. To estimate $\delta\beta$, we use the Landau energy in Eqs. (S18–S21) in the case $\mathbf{P} = 0$, $\boldsymbol{\phi}||[110]$ in the conjunction with Eq. (S24) and find $\delta\beta$ at which the

energy minimum coincides with the energy of the *Imcm* phase that we obtained from ab-initio. Using renormalized $\beta_{112}$, $\beta_{123}$ also affects the *Ima*2 phase (characterized by $\mathbf{P}||[110]$, $\boldsymbol{\phi}||[110]$), bringing its energy closer to the DFT-calculated counterpart (see Table S-II). Table S-I lists renormalized values of $\beta_{112}$, $\beta_{123}$, while the values $\beta_{112}^{\dagger}$, $\beta_{123}^{\dagger}$ are also listed in Table S-II. For all Landau-theory-based calculations, we use the renormalized values of $\beta_{112}$, $\beta_{123}$.

## Supplemental Material 4. Landau energy: modulated phases

To describe the modulated phases (such as *Pbam*), we introduce the following terms in addition to $F_{\mathrm{hom}}(\mathbf{P}, \boldsymbol{\phi})$ (in rotated coordinate system with $x$ along [110] and $y$ along [1$\bar{1}$0]):

$$
\begin{aligned}
F = F_{\mathrm{hom}}(\mathbf{P}, \boldsymbol{\phi}) &+ \frac{1}{2}G\Big(\frac{\partial P_y}{\partial x}\Big)^2 + \frac{1}{2}D_{11}\Big(\frac{\partial \phi_x}{\partial x}\Big)^2 + \frac{1}{2}D_{66}\Big(\frac{\partial \phi_y}{\partial x}\Big)^2 \\
&- WP_y\phi_y\frac{\partial \phi_x}{\partial x} - fP_y\frac{\partial \varepsilon_6}{\partial x} + \frac{1}{2}C\varepsilon_6^2 - R\phi_x\phi_y\varepsilon_6,
\end{aligned}
\tag{S25}
$$

where $\varepsilon_6 = \partial u_y/\partial x$ is the elastic shear strain, $u_y$ is the displacement field in $y$-direction, $G$, $D_{11}$, $D_{66}$ are the correlation coefficients for, respectively, polarization, longitudinal and transverse tilts, $W$ and $f$ are rotopolar and flexoelectric coupling, $C$ is the elastic shear modulus, and $R$ is the shear rotostriction. The values for these coefficients extracted from first principles are listed in Tab. S-III.

The energy density $F$ shown in Eq. (S25) reaches its minimum when the Euler-Lagrange equations for the Landau potential are satisfied, i.e., when the variational derivatives of $F$ over each order parameter ($P_y$, $\phi_y$, $\phi_x$, $\varepsilon_6$) are zero. Let us start with the Euler–Lagrange equation for $\varepsilon_6$:

$$
\frac{\delta F}{\delta \varepsilon_6} = C\varepsilon_6 - R\phi_x\phi_y + f\frac{\partial P_y}{\partial x} = 0 \quad \Rightarrow \quad \varepsilon_6 = \frac{R}{C}\phi_x\phi_y - \frac{f}{C}\frac{\partial P_y}{\partial x}.
\tag{S26}
$$

Substituting the solution for $\varepsilon_6$ back into Eq. (S25), we can rewrite the terms with $\varepsilon_6$ as follows (assuming negligible effect of $\partial \phi_y/\partial x$):

$$
\begin{aligned}
-fP_y\frac{\partial \varepsilon_6}{\partial x} &+ \frac{1}{2}C\varepsilon_6^2 - R\phi_x\phi_y\varepsilon_6 \\
&= -\frac{fR}{C}P_y\phi_y\frac{\partial \phi_x}{\partial x} + \frac{f^2}{2C}\Big(\frac{\partial P_y}{\partial x}\Big)^2 + \frac{f^2}{C}P_y\frac{\partial^2 P_y}{\partial x^2} - \frac{R^2}{2C}\phi_x^2\phi_y^2.
\end{aligned}
\tag{S27}
$$

| $G$ | $0.573 \times 10^{-3}$ | $W$ | $4.13 \times 10^{-4}$ | $C$ | $4.34 \times 10^{-3}$ |
|---|---|---|---|---|---|
| $D_{11}$ | $0.56 \times 10^{-3}$ | $f$ | $0.097 \times 10^{-3}$ | $R$ | $2.0 \times 10^{-4}$ |
| $D_{66}$ | $1.27 \times 10^{-3}$ | | | | |

TABLE S-III. Tensor components in Eq. (3) of the main text responsible for the *Pbam* phase. The values are given in the *rotated* coordinate system, where indices 1,2,3 correspond to [110], [1$\bar{1}$0], [001] pseudocubic crystallographic directions. All values shown are in atomic units (Ha, bohr); both polarization $\mathbf{P}$ and tilts $\boldsymbol{\phi}$ are measured in bohrs.

Integrating $F$ over a period of modulations results in $\int P_y(\partial^2 P_y/\partial x^2)dx = -\int (\partial P_y/\partial x)^2 dx$. This allows us to rewrite Eq. (S25) as follows:

$$F = F_{\text{hom}}^*(\mathbf{P}, \boldsymbol{\phi}) + \frac{1}{2}G^*\left(\frac{\partial P_y}{\partial x}\right)^2 + \frac{1}{2}D_{11}\left(\frac{\partial \phi_x}{\partial x}\right)^2 + \frac{1}{2}D_{66}\left(\frac{\partial \phi_y}{\partial x}\right)^2 - W^*P_y\phi_y\frac{\partial \phi_x}{\partial x}, \quad \text{(S28)}$$

$$F_{\text{hom}}^*(\mathbf{P}, \boldsymbol{\phi}) = F_{\text{hom}}(\mathbf{P}, \boldsymbol{\phi}) - \frac{R^2}{2C}\phi_x^2\phi_y^2, \qquad G^* = G - \frac{f^2}{C}, \qquad W^* = W + \frac{fR}{C}, \quad \text{(S29)}$$

recovering Eq. (4) of the main text.

The remaining Euler–Lagrange equations ($\delta F/\delta P_y = 0$, $\delta F/\delta \phi_y = 0$, $\delta F/\delta \phi_x = 0$) are solved numerically using COMSOL Multiphysics software with imposed periodic boundary conditions; the total energy of the system $\int F dx$ per formula unit is then analysed at varying periodicity in order to obtain the optimal period of modulations.

## Supplemental Material 5.   Anisotropy of $G^*$, $W^*$ in PbZrO$_3$ and other perovskites

For description of anisotropy of $G^*$ and $W^*$ coefficients, we consider a class of modulations $P_{\hat{y}}(\hat{x})$, where $\hat{x}$ and $\hat{y}$ are indices denoting two mutually perpendicular directions that do not necessarily follow symmetric crystallographic directions. For such modulations, Eq. (S25) can be rewritten in the form:

$$
\begin{aligned}
F = {} & F_{\text{hom}}(\mathbf{P}, \boldsymbol{\phi}) + \frac{1}{2}\hat{G}\left(\frac{\partial P_{\hat{y}}}{\partial \hat{x}}\right)^2 + \frac{1}{2}\hat{D}_{11}\left(\frac{\partial \phi_{\hat{x}}}{\partial \hat{x}}\right)^2 + \frac{1}{2}\hat{D}_{66}\left(\frac{\partial \phi_{\hat{y}}}{\partial \hat{x}}\right)^2 \\
& - \hat{W}P_{\hat{y}}\phi_{\hat{y}}\frac{\partial \phi_{\hat{x}}}{\partial \hat{x}} - \hat{f}P_{\hat{y}}\frac{\partial \varepsilon_{\hat{6}}}{\partial \hat{x}} + \frac{1}{2}\hat{C}\varepsilon_{\hat{6}}^2 - \hat{R}\phi_{\hat{x}}\phi_{\hat{y}}\varepsilon_{\hat{6}} \\
= {} & \hat{F}_{\text{hom}}^*(\mathbf{P}, \boldsymbol{\phi}) + \frac{1}{2}\hat{G}^*\left(\frac{\partial P_{\hat{y}}}{\partial \hat{x}}\right)^2 + \frac{1}{2}\hat{D}_{11}\left(\frac{\partial \phi_{\hat{x}}}{\partial \hat{x}}\right)^2 - \hat{W}^*P_{\hat{y}}\phi_{\hat{y}}\frac{\partial \phi_{\hat{x}}}{\partial \hat{x}},
\end{aligned}
\quad \text{(S30)}
$$

$$\hat{F}^*_{\mathrm{hom}}(\mathbf{P}, \boldsymbol{\phi}) = F_{\mathrm{hom}}(\mathbf{P}, \boldsymbol{\phi}) - \frac{\hat{R}^2}{2\hat{C}}\phi^2_{\hat{x}}\phi^2_{\hat{y}}, \qquad \hat{G}^* = \hat{G} - \frac{\hat{f}^2}{\hat{C}}, \qquad \hat{W}^* = \hat{W} + \frac{\hat{f}\hat{R}}{\hat{C}}, \qquad \text{(S31)}$$

where $\hat{G}$, $\hat{D}$, $\hat{W}$, $\hat{f}$, $\hat{C}$, $\hat{R}$ are the rotated tensor components corresponding to the coordinate framework $(\hat{x}, \hat{y}, \hat{z})$:

$$\hat{T}_{ijkl} = \sum_{\alpha\beta\gamma\delta} a_{i\alpha}a_{j\beta}a_{k\gamma}a_{l\delta}T_{\alpha\beta\gamma\delta}. \qquad \text{(S32)}$$

We use the following rotation matrix $a_{ij}$ describing the transformation between coordinates $(x_1, x_2, x_3)$ corresponding to [100], [010], [001] crystallographic directions and $(\hat{x}, \hat{y}, \hat{z})$:

$$a_{ij} = \begin{pmatrix} \cos\theta\cos\phi & \cos\theta\sin\phi & \sin\theta \\ -\sin\phi\cos t + \sin\theta\cos\phi\sin t & \cos\phi\cos t + \sin\theta\sin\phi\sin t & -\cos\theta\sin t \\ -\sin\phi\sin t - \sin\theta\cos\phi\cos t & \cos\phi\sin t - \sin\theta\sin\phi\cos t & \cos\theta\cos t \end{pmatrix}, \qquad \text{(S33)}$$

where three angles $\theta$, $\phi$ and $t$ uniquely describe the rotated coordinate system.

It is possible to show that the tensor components $\hat{G}$, $\hat{W}$, $\hat{f}$, $\hat{C}$ and $\hat{R}$ that appear in Eq. (S30) can be all written in the following form (with only two independent tensor components):

$$\hat{T}(\theta, \phi, t) = c(\theta, \phi, t)\hat{T}(0, 0, 0) + [1 - c(\theta, \phi, t)]\hat{T}(0, \pi/4, 0), \qquad 0 \le c(\theta, \phi, t) \le 1. \qquad \text{(S34)}$$

This means that for any combination of angles $\theta$, $\phi$, $t$ the tensor component $\hat{T}(\theta, \phi, t)$ always lies between the following two values: $\hat{T}(0, 0, 0)$ describing the tensor component in the coordinate frame $\hat{x}||[100]$, $\hat{y}||[010]$, $\hat{z}||[001]$, and $\hat{T}(0, \pi/4, 0)$ describing the tensor component in the coordinate frame $\hat{x}||[110]$, $\hat{y}||[1\bar{1}0]$, $\hat{z}||[001]$. In the following, we designate the two extremities as $T(\langle 100\rangle)$ and $T(\langle 110\rangle)$, respectively.

In other words, coefficients $G(\langle 110\rangle)$, $W(\langle 110\rangle)$, $f(\langle 110\rangle)$, $C(\langle 110\rangle)$, $R(\langle 110\rangle)$ appearing in Eq. (S25) for the *Pbam* geometry are all guaranteed to be either the absolute maxima or the absolute minima of their respective anisotropic angular-dependent tensor components, while the other extremity is given by $G(\langle 100\rangle)$, $W(\langle 100\rangle)$, $f(\langle 100\rangle)$, $C(\langle 100\rangle)$, $R(\langle 100\rangle)$. For this reason, listing the tensor components obtained for two modulation directions, $\langle 100\rangle$ and $\langle 110\rangle$ is sufficient for the analysis of the angular anisotropy of the tensor component. The corresponding tensor components for $PbZrO_3$, as compared with $BaTiO_3$ (data taken from Ref. 8) and $SrTiO_3$ [4, 5], are shown in Tab. S-IV. We also list coefficients $G^*$ and $W^*$ in Tab. S-IV for completeness.

| | PbZrO$_3$ | SrTiO$_3$ | BaTiO$_3$ |
|---|---|---|---|
| $G(\langle 100 \rangle)$ | $3.15 \times 10^{-3}$ | $3.87 \times 10^{-3}$ | $2.32 \times 10^{-3}$ |
| $G(\langle 110 \rangle)$ | $0.573 \times 10^{-3}$ | $13.8 \times 10^{-3}$ | $21.5 \times 10^{-3}$ |
| $f(\langle 100 \rangle)$ | $-0.387 \times 10^{-3}$ | $-1.05 \times 10^{-3}$ | $-0.640 \times 10^{-3}$ |
| $f(\langle 110 \rangle)$ | $0.0969 \times 10^{-3}$ | $0.684 \times 10^{-3}$ | $0.0502 \times 10^{-3}$ |
| $C(\langle 100 \rangle)$ | $2.10 \times 10^{-3}$ | $4.16 \times 10^{-3}$ | $4.63 \times 10^{-3}$ |
| $C(\langle 110 \rangle)$ | $4.33 \times 10^{-3}$ | $4.65 \times 10^{-3}$ | $3.89 \times 10^{-3}$ |
| $G^*(\langle 100 \rangle)$ | $3.08 \times 10^{-3}$ | $3.61 \times 10^{-3}$ | $2.23 \times 10^{-3}$ |
| $G^*(\langle 110 \rangle)$ | $0.571 \times 10^{-3}$ | $13.7 \times 10^{-3}$ | $21.5 \times 10^{-3}$ |
| $W(\langle 100 \rangle)$ | $-2.0 \times 10^{-4}$ | $4.1 \times 10^{-4}$ | $-$ |
| $W(\langle 110 \rangle)$ | $4.1 \times 10^{-4}$ | $-3.0 \times 10^{-4}$ | $-$ |
| $R(\langle 100 \rangle)$ | $-0.45 \times 10^{-4}$ | $-2.4 \times 10^{-4}$ | $-$ |
| $R(\langle 110 \rangle)$ | $2.0 \times 10^{-4}$ | $4.5 \times 10^{-4}$ | $-$ |
| $W^*(\langle 100 \rangle)$ | $-1.92 \times 10^{-4}$ | $4.7 \times 10^{-4}$ | $-$ |
| $W^*(\langle 110 \rangle)$ | $4.14 \times 10^{-4}$ | $-2.9 \times 10^{-4}$ | $-$ |

TABLE S-IV. Anisotropy of tensor components, in Hartree atomic units.

## Supplemental Material 6.   Analytical expression for the *Ima2*–*Pbam* phase diagram

The necessary condition for the modulations to exist in *Pbam*-like phases of PZO is that their energy be lower than the energy of their non-modulated counterpart, i.e., of the *Ima2* phase. This is equivalent to the problem of finding at what material parameters the energy of a FE domain wall in the *Ima2* structure is negative.

Let us consider a FE domain wall in the *Ima2* structure, at which $P_y(x)$ changes its sign. At such a wall the $y$-component of the tilt can be considered constant ($\phi_y = \phi_{y0}$), while $\phi_x(x) \neq 0$ is produced at the wall via the rotopolar coupling to the polarization gradient. The energy of such a domain wall is calculated by integrating the excess energy $F$ in Eq. (S28) over the energy of *Ima2* phase:

$$E_{\mathrm{DW}}\Big[P_y(x), \phi_{y0}, \phi_x(x)\Big] = \int_{-\infty}^{\infty} (F - E_{Ima2}) dx. \tag{S35}$$

Distributions $P_y(x)$, $\phi_x(x)$ at the wall are obtained by minimizing $E_{\text{DW}}$, i.e., by solving Euler-Lagrange equations for variational derivatives, $\dfrac{\delta F}{\delta P_y} = \dfrac{\delta F}{\delta \phi_x} = 0$.

Let $(G^*, W^*)$ be a pair of the material parameters lying on the $Pbam$–$Ima2$ phase transition contour, resulting in zero energy of the wall. When varying parameters $G^*$, $W^*$ by some $dG^*$, $dW^*$, the spatial distributions of the order parameters at the wall $P_y(x)$, $\phi_x(x)$ will change by some $\delta P_y(x)$, $\delta \phi_x(x)$, leading to the following change in the domain wall energy:

$$
\begin{aligned}
dE_{\text{DW}} &= \int_{-\infty}^{\infty} \left[ \frac{\delta F}{\delta P_y} \delta P_y + \frac{\delta F}{\delta \phi_x} \delta \phi_x + \frac{\partial F}{\partial W^*} dW^* + \frac{\partial F}{\partial G^*} dG^* \right] dx \\
&= \int_{-\infty}^{\infty} \left[ \phi_{y0} \phi_x \frac{dP_y}{dx} dW^* + \frac{1}{2} \left( \frac{dP_y}{dx} \right)^2 dG^* \right] dx,
\end{aligned}
\tag{S36}
$$

where we used $\dfrac{\delta F}{\delta P_y} = \dfrac{\delta F}{\delta \phi_x} = 0$. A point $G^* + dG^*, W^* + dW^*$ lies on the same phase transition contour if the energy of the domain wall remains unchanged, $dE_{\text{DW}} = 0$, leading to the following condition on $dG^*, dW^*$:

$$
\frac{dW^*}{dG^*} = -\frac{\int_{-\infty}^{\infty} \frac{1}{2} (dP_y/dx)^2 dx}{\int_{-\infty}^{\infty} \phi_{y0} \phi_x (dP_y/dx) dx}.
\tag{S37}
$$

In order to calculate integrals in Eq. (S37), one needs to know the spatial distributions of $P_y(x)$ and $\phi_x(x)$ at the wall, which we are going to estimate. The leading orders of the equations for $P_y(x)$ and $\phi_x(x)$, obtained from $\dfrac{\delta F}{\delta P_y} = \dfrac{\delta F}{\delta \phi_x} = 0$, are:

$$
\alpha_1^* P_y + \alpha_{11}^* P_y^3 = G^* \frac{d^2 P_y}{dx^2},
\tag{S38}
$$

$$
\beta_1^* \phi_x = D_{11} \frac{d^2 \phi_x}{dx^2} - W^* \phi_{y0} \frac{dP_y}{dx},
\tag{S39}
$$

$$
\alpha_1^* = \frac{\partial^2 F_{\text{hom}}^*}{\partial P_y^2} \left( 0, \phi_{y0} \right) = \alpha_1 - 2Q_{11}\phi_{y0}^2 + 2T_{111}\phi_{y0}^4 < 0,
\tag{S40}
$$

$$
\alpha_{11}^* = \alpha_{11} + 4S_{111}\phi_{y0}^2 > 0,
\tag{S41}
$$

$$
\begin{aligned}
\beta_1^* = \frac{\partial^2 F_{\text{hom}}^*}{\partial \phi_x^2} \left( 0, \phi_{y0} \right) &= \beta_1 + \left( \beta_{12} - \frac{R^2}{C} \right) \phi_{y0}^2 \\
&+ 2\beta_{112}\phi_{y0}^4 + 2\beta_{1112}\phi_{y0}^6 - 2Q_{12}P_{y0}^2 + 2S_{112}P_{y0}^4 + 2T_{123}P_{y0}^2\phi_{y0}^2 > 0,
\end{aligned}
\tag{S42}
$$

where $P_{y0}$ is the spontaneous polarization in the $Ima2$ phase. We neglect the effect of the rotopolar term in Eq. (S38), assuming that the polarization correlation length $\xi_P = \sqrt{G^*/|\alpha_1^*|}$ is smaller than the longitudinal tilt correlation length $\xi_\phi = \sqrt{D_{11}/\beta_1^*}$.

The solution to Eq. (S38) is

$$P_y(x) = P_{y0} \tanh\left(\frac{x}{\sqrt{2}\xi_P}\right),\tag{S43}$$

$$\int_{-\infty}^{\infty} \frac{1}{2}\left(\frac{dP_y}{dx}\right)^2 dx = \frac{\sqrt{2}P_{y0}^2}{3\xi_P}.\tag{S44}$$

To estimate the solution to Eq. (S39), we approximate $P_y(x)$ with a piece-wise function:

$$P_y(x) \approx \begin{cases} -P_{y0}, & x < -\sqrt{2}\xi_P, \\ \frac{P_{y0}}{\sqrt{2}\xi_P}x, & -\sqrt{2}\xi_P < x < \sqrt{2}\xi_P, \\ P_{y0}, & x > \sqrt{2}\xi_P, \end{cases}\tag{S45}$$

This approximation allows us to rewrite Eq. (S39) as follows:

$$\frac{d^2\phi_x}{dx^2} - \frac{1}{\xi_\phi^2}\phi_x = \begin{cases} \frac{W^*\phi_{y0}P_{y0}}{\sqrt{2}\xi_P D_{11}}, & -\sqrt{2}\xi_P < x < \sqrt{2}\xi_P, \\ 0, & \text{otherwise}, \end{cases}\tag{S46}$$

solution to which is

$$\phi_x(x) = \begin{cases} -\frac{W^*\phi_{y0}P_{y0}\xi_\phi^2}{\sqrt{2}\xi_P D_{11}}\left[1 - e^{-\sqrt{2}\xi_P/\xi_\phi}\cosh(x/\xi_\phi)\right], & -\sqrt{2}\xi_P < x < \sqrt{2}\xi_P, \\ -\frac{W^*\phi_{y0}P_{y0}\xi_\phi^2}{\sqrt{2}\xi_P D_{11}}\sinh(\sqrt{2}\xi_P/\xi_\phi)e^{-|x|/\xi_\phi}, & \text{otherwise}, \end{cases}\tag{S47}$$

$$\int_{-\infty}^{\infty} \phi_{y0}\phi_x\frac{dP_y}{dx}dx = -\frac{2W^*\phi_{y0}^2 P_{y0}^2\xi_\phi}{D_{11}} + o(\xi_P/\xi_\phi),\tag{S48}$$

where $\xi_P \ll \xi_\phi$ is assumed.

Using Eqs. (S44,S48), we can rewrite the condition for $dW^*/dG^*$ as follows:

$$\frac{dW^*}{dG^*} = \frac{\sqrt{2|\alpha_1^*|\beta_1^* D_{11}}}{6\phi_{y0}^2}\frac{1}{\sqrt{G^*W^*}},\tag{S49}$$

solution to which is

$$W^* = \frac{1}{\phi_{y0}}\left(\frac{8}{9}|\alpha_1^*|\beta_1^* D_{11}G^*\right)^{1/4},\tag{S50}$$

where the trivial point $(G^*, W^*) = (0,0)$ lying at the phase transition contour was used. This corresponds to Eq. (5) of the main text, where $\gamma = \frac{9}{8|\alpha_1^*|\beta_1^*}$.

We would like to emphasize that we did not use any empirical fitting to obtain this analytical expression.

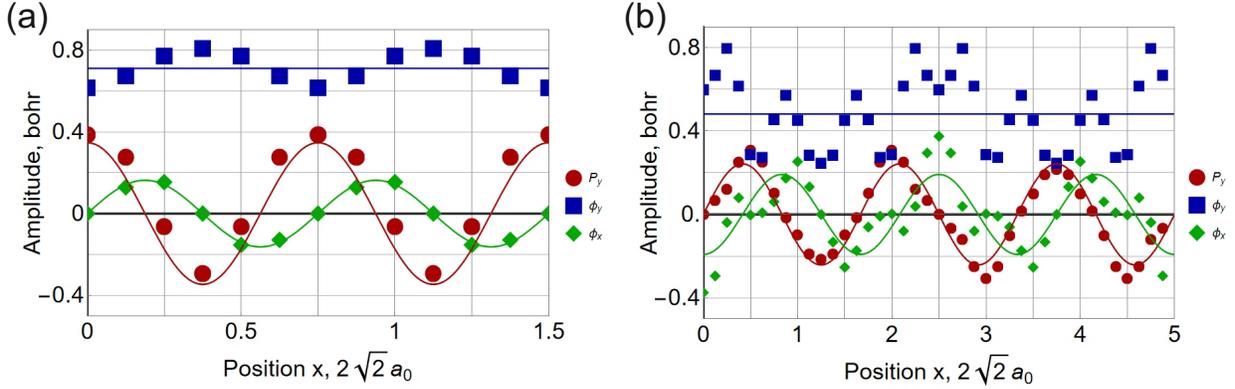

FIG. S2. Layer-by-layer decomposition of (a) ab-initio ↑↑↓ FiE structure of PbZrO$_3$, (b) X-ray-observed IC structure of PbHfO$_3$.

## Supplemental Material 7.   Examples of modulated phases in perovskites: ferrielectric (ab initio) and real incommensurate (X-ray)

Figure S2 (a) shows the layer-by-layer decomposition of the ferrielectric ↑↑↓ structure that we obtain from ab-initio, analogous to Fig. 1(b) of the main text. The main features are the same as in the AFE ↑↑↓↓ structure: $P_y$ and $\phi_x$ are modulated and phase-shifted by 90°, while $\phi_y$ remains mostly constant across the structure (though there are also small modulations of $\phi_y$ with amplitude $\sim 15\%$ of the mean value).

We have also analysed the X-ray data of the IC phase of PbHfO$_3$ reported in e.g. [9], in which the observed modulation period is 67% larger than in the *Pbam* phase of PZO. Based on the atomic positions listed in that work we extract the spatial distributions of the polarization and of the AFD tilts and plot them in Fig. S2(b). The obtained order parameters decomposition of the real IC structure is close to our predictions [cf. Fig. 1(b) and Fig. S2(a)]: primary AFD tilt component $\phi_y$ is relatively uniform over the modulation period (with an additional double-wavevector modulation consistent with the biquadratic coupling between $P$ and $\phi$) and, most importantly, secondary tilt $\phi_x$ has a well-defined non-zero amplitude and phase shifted by $\pi/2$ with respect to the polarization. This observation unambiguously confirms the relevance of the rotopolar coupling in polar IC structures in perovskites in general.



\end{document}